\documentclass[12pt]{article}
\usepackage{graphicx}

\newcommand\pubnumber{WSU-HEP-1004}
\newcommand\pubdate{\today}

\def\wsu{Department of Physics and Astronomy\\
Wayne State University, Detroit, MI 48201, USA}
\def\support{\footnote{This work was supported in part by the U.S.\ National Science Foundation
CAREER Award PHY--0547794, and by the U.S.\ Department of Energy under Contract
DE-FG02-96ER41005.}}

\def\Title#1{\begin{center} {\Large #1 } \end{center}}
\def\Author#1{\begin{center}{ \sc #1} \end{center}}
\def\Address#1{\begin{center}{ \it #1} \end{center}}

\newcommand\pubblock{\rightline{\begin{tabular}{l} \pubnumber\\
         \pubdate  \end{tabular}}}
\newenvironment{Abstract}{\begin{quotation}  }{\end{quotation}}
\newenvironment{Presented}{\begin{quotation} \begin{center} 
             PRESENTED AT\end{center}\bigskip 
      \begin{center}\begin{large}}{\end{large}\end{center} \end{quotation}}

\def\beq{\begin{equation}}
\def\eeq#1{\label{#1}\end{equation}}
\def\eeqn{\end{equation}}

\def\beqa{\begin{eqnarray}}
\def\eeqa#1{\label{#1}\end{eqnarray}}
\def\eeqan{\end{eqnarray}}

\let\bar=\overbar

\def\D{{\cal D}}

\def\Dslash{\not{\hbox{\kern-4pt $D$}}}
\def\dslash{\not{\hbox{\kern-2pt $\del$}}}

\def\msb{{\bar{\ssstyle M \kern -1pt S}}}

\def\barD{\overline D{}^0}
\def\D0bar{\overline D{}^0}
\def\DDbar{D{}^0-\overline D{}^0}

\def\beq{\begin{equation}}
\def\eeq{\end{equation}}
\def\bea{\begin{eqnarray}}
\def\eea{\end{eqnarray}}

\newcommand{\Dzb}{\overline{D^0}}
\newcommand{\DzDzb}{D^0-\overline{D^0}}

\begin{document}
\begin{titlepage}
\pubblock

\vfill
\Title{CP-violation in charm}
\vfill
\Author{ Alexey A Petrov\support}
\Address{\wsu}
\vfill
\begin{Abstract}
I review recent results in theoretical and experimental analyses of CP-violation in charmed transitions, paying
particular attention to constraints on parameters of beyond the Standard Model interactions. 
\end{Abstract}
\vfill
\begin{Presented}
The 6th International Workshop on the 
CKM Unitarity Triangle (CKM2010), University of Warwick, UK, 6-10 September 2010
\end{Presented}
\vfill
\end{titlepage}
\def\thefootnote{\fnsymbol{footnote}}
\setcounter{footnote}{0}
%

\section{Introduction}

Charm transitions play an important role in flavor physics. Along with the corresponding searches in 
strange and beauty-flavored systems, charm provide outstanding opportunities for 
indirect searches for physics beyond the Standard Model (SM). These searches yield stringent 
constraints on the models of New Physics (NP) because of the availability of large statistical samples of 
charm data~\cite{Artuso:2008vf}.
 
One of the most important tools in indirect studies of New Physics is the observation of CP-violation. 
The Standard Model's picture of CP-violation~\cite{BigiSandaBook} is related to the phases of the  
coupling constants of dimension-four operators describing quark Yukawa interactions with Higgs fields $\phi$,
\beq
{\cal L}_Y = \xi_{ik} \overline \psi_i \psi_k \phi + \mbox{~h.c.}
\eeq
These complex Yukawa couplings $\xi_{ik}$ lead to a complex-valued 
Cabibbo-Kobayashi-Maskawa (CKM) quark mixing matrix providing a
natural source of CP-violation for the case of the Standard Model with three 
(or more) generations. The SM with three generations has a single CP-violating 
phase, making it a very restrictive system with a possibility to
relate observed effects in quark systems with different flavors. This mechanism 
was experimentally confirmed in the observations of oscillations and decays of  
beauty and strange mesons.

This is clearly not a unique way of introducing CP-violation in Quantum Field Theory. 
Another way involves adding operators of dimensions less than 
four (the ``soft'' CP-breaking), which is popular in supersymmetric models.
Yet another way is to break CP-invariance spontaneously. This method, which 
is somewhat aesthetically appealing, introduces a CP-violating ground state with
a CP-conserved Lagrangian. It is realized in a class of left-right-symmetric models 
or multi-Higgs models. All these mechanisms can be probed in charm transitions.

It can be argued that the observation of CP-violation in the current round of charm experiments
constitutes one of the signals of physics beyond the Standard Model (BSM).
This argument stems from the fact that all quarks that build up initial and final 
hadronic states in weak decays of charm mesons or baryons belong to the first two generations. 
This implies that those transitions are governed by a $2\times2$ Cabibbo quark mixing matrix.
This matrix is real, so no CP-violation is possible in the
dominant tree-level diagrams which describe the decay amplitudes. 
In the Standard Model, CP-violating amplitudes in charm transitions can be introduced 
by including penguin or box operators induced by virtual $b$-quarks. However, their 
contributions are strongly suppressed by the small combination of 
CKM matrix elements $V_{cb}V^*_{ub}$. Explicit evaluations of $b$-quark contributions 
to mixing asymmetries yield results of the order of $0.1-1\%$~\cite{Bobrowski:2010xg} with similar
predictions for decay amplitudes~\cite{Buccella:1992sg}.
Thus, observation of larger CP violation in charm decays or mixing would be an 
unambiguous sign for new physics.

As with other flavor physics, CP-violating contributions in charm can be generally 
classified by three different categories:

\begin{enumerate}
\item[(I)] 
CP violation in the $\DzDzb$ mixing matrix (or ``indirect" CP-violation). Introduction of 
$\Delta C = 2$ transitions, either via SM or NP one-loop or tree-level NP 
amplitudes leads to non-diagonal entries in the $D^0-\barD$ mass matrix,
\beq\label{MixingMatrix}
\left[M - i \frac{\Gamma}{2} \right]_{ij} = 
\left(
\begin{array}{cc}
A & p^2 \\
q^2 & A 
\end{array} 
\right)
\eeq
This type of CP violation is manifest when 
$R_m^2=\left|p/q\right|^2=(2 M_{12}-i \Gamma_{12})/(2 M_{12}^*-i 
\Gamma_{12}^*) \neq 1$.

\item[(II)]
CP violation in the $\Delta C =1$ decay amplitudes (or ``direct'' CP-violation). 
This type of CP violation 
occurs when the absolute value of the decay amplitude for $D$ to decay to a 
final state $f$ ($A_f$) is different from the one of the corresponding 
CP-conjugated amplitude (``direct CP-violation''). This can happen if
the decay amplitude can be broken into at least two parts associated with 
different weak and strong phases,
\beq\label{DirectAmpl}
A_f =
\left|A_1\right| e^{i \delta_1} e^{i \phi_1} +
\left|A_2\right| e^{i \delta_2} e^{i \phi_2},
\eeq
where $\phi_i$ represent weak phases ($\phi_i \to -\phi_i$ under CP-transormation),
and $\delta_i$ represents strong phases ($\delta_i \to \delta_i$ under CP-transformation).
This ensures that the CP-conjugated amplitude, $\overline A_{\overline f}$ would differ 
from $A_f$.

\item[(III)] CP violation in the interference of decays with and without mixing.
This type of CP violation is possible for a subset of final states to which
both $D^0$ and $\Dzb$ can decay. 
\end{enumerate}
For a given final state $f$, CP violating contributions can be summarized 
in the parameter 
\begin{equation}\label{Lambda}
\lambda_f = \frac{q}{p} \frac{{\overline A}_f}{A_f}=
R_m e^{i(\phi+\delta)}\left| \frac{{\overline A}_f}{A_f}\right|,
\end{equation}
where $A_f$ and ${\overline A}_f$ are the amplitudes for $D^0 \to f$ and 
$\Dzb \to f$ transitions respectively and $\delta$ is the CP-conserving strong phase 
difference between $A_f$ and ${\overline A}_f$. In Eq.~(\ref{Lambda}) $\phi$ represents the
convention-independent CP-violating phase difference between the ratio of 
decay amplitudes and the mixing matrix.

\section{Indirect CP-violation}

The non-diagonal entries in the mixing matrix of Eq.~(\ref{MixingMatrix})
lead to mass eigenstates of neutral $D$-mesons that are different from 
the weak eigenstates. They, however, are related by a linear transformation,
\beq
|D_{1\atop 2} \rangle = p | D^0 \rangle \pm q | \overline{D}^0 \rangle,
\eeq
where the complex parameters $p$ and $q$ are obtained from diagonalizing the 
$\DDbar$ mass matrix of Eq.~(\ref{MixingMatrix}). Note that if CP-violation is neglected, 
$p=q=1/\sqrt{2}$. The mass and width splittings between mass eigenstates are 
\beq\label{XandY}
x_D= \frac{m_1-m_2}{\Gamma_D}, \qquad y_D=\frac{\Gamma_1-\Gamma_2}{2 \Gamma_D},
\eeq
where $\Gamma_{\rm D}$ is the average width of the two neutral $D$ meson mass eigenstates.  
Because of the absence of superheavy down-type quarks destroying Glashow-Iliopoulos-Maiani 
(GIM) cancellation, it is expected that $x_D$ and $y_D$ should be rather small in the Standard Model.
The quantities which are actually measured in experimental determinations of the mass and width differences, 
are $y_{\rm D}^{\rm (CP)}$ (measured in time-dependent $D \to KK, \pi\pi$ analyses), 
$x_{\rm D}'$, and $y_{\rm D}'$ (measured, e.g., in $D \to K\pi$ or similar transitions), are defined as
\bea
y_{\rm D}^{\rm (CP)} &=&  y_{\rm D} \cos\phi - x_{\rm D}\sin\phi
\left(\frac{A_m}{2}-A_{prod}\right) \ \ , \nonumber \\
x_D' &=& x_D\cos\delta_{K\pi} + y_D\sin\delta_{K\pi} \ \ ,
\\
y_D' &=& y_{\rm D} \cos \delta_{K\pi} - x_{\rm D}
\sin\delta_{K\pi} \ \ ,
\nonumber
\label{y-defs}
\eea
where
$A_{prod} = \left(N_{D^0} - N_{{\overline D}^0}\right)/
\left(N_{D^0} + N_{{\overline D}^0}\right)$ is the so-called
production asymmetry of $D^0$ and $\overline{D}^0$ (giving
the relative weight of $D^0$ and ${\overline D}^0$ in the
sample) and $\delta_{K\pi}$ is the strong phase difference between
the Cabibbo favored and double Cabibbo suppressed
amplitudes~\cite{Bergmann:2000id}, which  can be measured in 
$D\to K\pi$ transitions. A CP-violating phase $\phi$ is defined in Eq.~(\ref{Lambda}).
A fit to the current database of experimental analyses 
by the Heavy Flavor Averaging Group (HFAG) gives~\cite{ExperimentalAnalyses,HFAG}
\bea\label{hfag}
& & x_{\rm D} = 0.0100^{+0.0024}_{-0.0026}~, ~~\qquad 
y_{\rm D} = 0.0076^{+0.0017}_{-0.0018} \  \ ,
\nonumber \\
& & 1 - |q/p| = 0.06 \pm 0.14, \quad 
\phi = -0.05 \pm 0.09.
\eea
At this stage it is important to note that the size of the signal allows to conclude that the former 
"smoking gun" signal for New Physics in $\DDbar$ mixing, $x \gg y$ no longer applies. 
Now, even though theoretical calculations of $x_D$ and $y_D$ are quite 
uncertain, the values $x_D \sim y_D \sim 1\%$ are natural in the Standard Model~\cite{Falk:2001hx}. 
Also, as was argued earlier, CP-violation asymmetries in charm mixing are quite small. 
The question that arises now is how to use the data provided by 
Eq.~(\ref{hfag}) to probe physics beyond the SM.

This question can be answered using an effective field theory approach. Heavy BSM degrees of freedom cannot be 
directly produced in charm meson decays, but can nevertheless affect the effective $|\Delta C| = 2$ Hamiltonian by 
changing Wilson coefficients and/or introducing new operator structures\footnote{NP can also affect 
$|\Delta C| = 1$ transitions and thus contribute to $y_D$. For more details, see~\cite{Golowich:2006gq}.}. 
By integrating out those new degrees of freedom associated with new interactions at a 
high scale $M$, we are left with an effective hamiltonian written in the form of a series of operators of increasing 
dimension. It turns out that a model-independent study of NP $|\Delta C| = 2$ contributions is possible, as 
any NP model will only modify Wilson coefficients of those operators~\cite{Gedalia:2009kh,Golowich:2007ka},
\beq\label{SeriesOfOperators}
{\cal H}_{NP}^{|\Delta C| = 2}  =\frac{1}{M^2} \left[ 
\sum_{i=1}^8  {\rm C}_i (\mu) ~ Q_i  \right],
\eeq
where ${\rm C}_i$ are dimensionless Wilson coefficients, and the $Q_i$ are the effective operators:
\beqa
\begin{array}{l}
Q_1 = (\overline{u}_L^\alpha \gamma_\mu c_L^\alpha) \ 
(\overline{u}_L^\beta \gamma^\mu c_L^\beta)\ , \\
Q_2 = (\overline{u}_R^\alpha c_L^\alpha) \ 
(\overline{u}_R^\beta c_L^\beta)\ , \\
Q_3 = (\overline{u}_R^\alpha c_L^\beta) \ 
(\overline{u}_R^\beta c_L^\alpha) \ , \\
Q_4 = (\overline{u}_R^\alpha c_L^\alpha) \ 
(\overline{u}_L^\beta c_R^\beta) \ ,
\end{array}
\qquad
\begin{array}{l}
Q_5 = (\overline{u}_R^\alpha c_L^\beta) \ 
(\overline{u}_L^\beta c_R^\alpha) \ , \\
Q_6 = (\overline{u}_R^\alpha \gamma_\mu c_R^\alpha) \ 
(\overline{u}_R^\beta \gamma^\mu c_R^\beta)\ , \\
Q_7 = (\overline{u}_L^\alpha c_R^\alpha) \ 
(\overline{u}_L^\beta c_R^\beta)\ , \\
Q_8 = (\overline{u}_L^\alpha c_R^\beta) \ 
(\overline{u}_L^\beta c_R^\alpha) \ \ ,
\end{array}
\label{SetOfOperators}
\eeqa
where $\alpha$ and $\beta$ are color indices. In total, there are eight possible operator structures that exhaust the
list of possible independent contributions to $|\Delta C|=2$ transitions\footnote{Note that earlier 
Ref.~\cite{Golowich:2007ka} used a slightly different set of operators than~\cite{Gedalia:2009kh}, 
which can be related to each other by a linear transformation.}. Taking operator mixing into account, 
a set of constraints on the Wilson coefficients of Eq.~(\ref{SeriesOfOperators}) can be placed,
\bea
\begin{array}{l}
\left| C_1 \right| \leq 5.7 \times 10^{-7} \left[\frac{M}{1~\mbox{TeV}} \right]^2, 
\\
\left| C_2 \right| \leq 1.6 \times 10^{-7} \left[\frac{M}{1~\mbox{TeV}} \right]^2, 
\\
\left| C_3 \right| \leq 5.8 \times 10^{-7} \left[\frac{M}{1~\mbox{TeV}} \right]^2, 
\end{array}
\qquad
\begin{array}{l}
\left| C_4 \right| \leq 5.6 \times 10^{-8} \left[\frac{M}{1~\mbox{TeV}} \right]^2, 
\\
\left| C_5 \right| \leq 1.6 \times 10^{-7} \left[\frac{M}{1~\mbox{TeV}} \right]^2.
\end{array}
\label{ConstraintsOnCoefficients}
\eea
The constraints on $C_6-C_8$ are identical to those on $C_1-C_3$~\cite{Gedalia:2009kh}.
Note that Eq.~(\ref{ConstraintsOnCoefficients}) implies that New Physics particles, for some
unknown reason, have highly suppressed couplings to charmed quarks. Alternatively, 
the tight constraints of Eq.~(\ref{ConstraintsOnCoefficients}) probes NP at very high scales:
$M \ge (4-10) \times 10^3$~TeV for tree-level NP-mediated charm mixing and 
$M \ge (1-3) \times 10^2$~TeV for loop-dominated mixing via New Physics particles.

No CP-violation has been observed in charm transitions yet. However, available 
experimental constraints of Eq.~(\ref{hfag}) can provide some tests of CP-violating NP models. 
For example, a set of constraints on the imaginary parts of Wilson coefficients of 
Eq.~(\ref{SeriesOfOperators}) can be placed,
\bea
\begin{array}{l}
\mbox{Im} \left[C_1\right] \leq 1.1 \times 10^{-7} \left[\frac{M}{1~\mbox{TeV}} \right]^2, 
\\
\mbox{Im} \left[C_2\right] \leq 2.9 \times 10^{-8} \left[\frac{M}{1~\mbox{TeV}} \right]^2, 
\\
\mbox{Im} \left[C_3\right]  \leq 1.1 \times 10^{-7} \left[\frac{M}{1~\mbox{TeV}} \right]^2, 
\end{array}
\qquad
\begin{array}{l}
\mbox{Im} \left[C_4\right]  \leq 1.1 \times 10^{-8} \left[\frac{M}{1~\mbox{TeV}} \right]^2, 
\\
\mbox{Im} \left[C_5\right]  \leq 3.0 \times 10^{-8} \left[\frac{M}{1~\mbox{TeV}} \right]^2.
\end{array}
\eea
Just like the constraints of Eq.~(\ref{ConstraintsOnCoefficients}), they give a sense of
how NP particles couple to the Standard Model.

Other tests can also be performed. For instance, neglecting direct CP-violation in the
decay amplitudes, one can write a "theory-independent" relation among
$\DDbar$ mixing amplitudes~\cite{Grossman:2009mn,Kagan:2009gb},
\beq
\frac{x}{y} = \frac{1-|q/p|}{\tan\phi} 
\eeq
Current experimental results $x/y \approx 0.8 \pm 0.3$ imply that amount of CP-violation in the 
$\DDbar$ mixing matrix is comparable to CP-violation in the interference of decays and mixing
amplitudes. 

\section{Direct CP-violation}

In principle, $\DDbar$ mixing is not required for the observation of CP-violation. While
CPT-symmetry requires the total widths of $D$ and $\overline D$ to be the same, the
partial decay widths $\Gamma(D \to f)$ and $\Gamma({\overline D} \to {\overline f})$
could be different in the presence of CP-violation, which would be signaled by 
a non-zero value of the asymmetry
\begin{eqnarray}\label{Acp}
a_f=\frac{\Gamma(D \to f)-\Gamma({\overline D} \to {\overline f})}{
\Gamma(D \to f)+\Gamma({\overline D} \to {\overline f})}.
\end{eqnarray}
One can also introduce a related asymmetry $a_{\overline f}$ by substituting 
$f\to \bar f$ in Eq.~(\ref{Acp}). For charged $D$-decays the only contribution to the asymmetry 
of Eq.~(\ref{Acp}) comes from the multi-component structure of the
$\Delta C =1$ decay amplitude of Eq.~(\ref{DirectAmpl}). In this case,
\begin{eqnarray}\label{DCPaF}
a_f &=& \frac{2 Im\left(A_1 A_2^*\right) \sin\delta}
{\left|A_1\right|^2 + \left|A_2\right|^2 + 2 Re A_1 A_2^* \cos\delta}
= 2 r_f \sin\phi \sin\delta,
\end{eqnarray}
where $\delta = \delta_1-\delta_2$ is the CP-conserving phase difference and 
$\phi$ is the CP-violating one. $r_f=|A_2/A_1|$ is the ratio of amplitudes.
Both $r_f$ and $\delta$ are extremely difficult to compute reliably in 
$D$-decays. However, the task can be significantly simplified if one
only concentrates on detection of New Physics in CP-violating asymmetries in the
current round of experiments~\cite{Grossman:2006jg}, i.e. at the 
${\cal O}(1\%)$ level. This is the level at which $a_f$ is currently probed
experimentally, see, e.g. \cite{D0pipi}.
As follows from Eq.~(\ref{DCPaF}), in this case one should expect $r_f \sim 0.01$. 

It is easy to see that the Standard Model asymmetries are safely below this estimate.
First, Cabibbo-favored ($A_f \sim \lambda^0$) and doubly Cabibbo-suppressed 
($A_f \sim \lambda^2$) decay modes proceed via amplitudes that share the
same weak phase, so no CP-asymmetry is generated\footnote{Technically, there is 
a small, ${\cal O}(\lambda^4)$ phase difference between the dominant tree $T$ amplitude
and exchange $E$ amplitudes.}. 
On the other hand, singly-Cabibbo-suppressed decays ($A_f \sim \lambda^1$) readily 
have a two-component structure, receiving contributions from both tree and penguin 
amplitudes. In this case the same conclusion follows from the consideration of 
the charm CKM unitarity,
$
V_{ud} V_{cd}^* + V_{us} V_{cs}^* + V_{ub} V_{cb}^* = 0.
$

In the Wolfenstein parameterization of CKM, the first two terms in this equation are of 
the order ${\cal O}(\lambda)$ (where $\lambda \simeq 0.22$), while the last one is
${\cal O}(\lambda^5)$. Thus, the CP-violating asymmetry is expected to be at most 
$a_f \sim 10^{-3}$ in the SM. Model-dependent estimates of this 
asymmetry exist and are consitent with this estimate~\cite{Buccella:1992sg}.

Asymmetries of Eq.~(\ref{Acp}) can also be introduced for the neutral $D$-mesons.
In this case a much richer structure becomes available due to interplay of
CP-violating contributions to decay and mixing amplitudes~\cite{Bergmann:2000id,Grossman:2006jg}, 
\begin{eqnarray}\label{AcpNeutral}
a_f &=& a_f^d + a_f^m + a_f^i,
\nonumber \\
a_f^d &=& 2 r_f \sin\phi \sin\delta,
\\
a_f^m &=& - R_f \frac{y'}{2} \left(R_m - R_m^{-1}\right) \cos\phi,
\nonumber \\
a_f^i &=& R_f \frac{x'}{2} \left(R_m + R_m^{-1}\right) \sin\phi,
\nonumber 
\end{eqnarray}
where $ a_f^d$, $a_f^m$, and $a_f^i$ represent CP-violating contributions
from decay, mixing and interference between decay and mixing amplitudes respectively.
For the final states that are also CP-eigenstates, $f = \overline f$ and $y' = y$. All those 
asymmetries can be studied experimentally.

\section{Conclusions and outlook}

Studies of CP-violation will help to distinguish among the models of New Physics 
describing new particles possibly observed at the Large Hadron Collider (LHC) in the 
upcoming years. Recent studies of charm quark observables already 
revealed puzzling non-universality of possible NP contributions to low energy flavor-changing
transitions. In particular, no new signals of CP-violation have been observed.

An extensive experimental study of exclusive decays should be 
performed~\cite{Grossman:2006jg}, shedding new light on how large
CP-violation in charm transition amplitudes could be. Finally, new observables, such as
CP-violating "untagged" decay asymmetries~\cite{Petrov:2004gs} should be studied in
hadronic decays of charmed mesons. These analyses will be be indispensable for 
physics of the LHC era.


\end{document}